\documentstyle[graphicx,sprocl]{article}

\DeclareGraphicsExtensions{.eps}

\def\GeV{\;\mbox{GeV}}

\def\D#1{\displaystyle{#1}}
\title{THE STRONG COUPLING AND THE GLUON DENSITY FROM JETS IN DIS} 
\author{T. HADIG}
\address{
Physikalisches Institut, Universit{\"a}t Heidelberg, 69120 Heidelberg, Germany\\ 
previously: I.\ Physikalisches Institut, RWTH Aachen, 52056 Aachen, Germany\\
on behalf of the H1 and ZEUS Collaborations}

\begin{document}
\maketitle

\abstracts{
The extraction of the parameters of perturbative QCD is one of the main 
tasks for the HERA experiments. Studies on the structure of the hadronic
final states with jet algorithms give a direct handle on the parton
density functions and the strong coupling strength over a wide kinematic
range. In this article, new results of the H1 and ZEUS collaborations
for the strong coupling and the gluon density in the proton are presented.
}

\section {Introduction}

The observation of the hadronic final state in lepton hadron
interactions is an essential tool for tests of QCD. A general feature of
the events is the non-isotropic energy flow, i.e.~large fractions of the
particles created in the interactions are concentrated in small regions
in space. Jet algorithms are used to measure the number and properties
of the energy clusters.

Theoretical calculations for the cross sections of these processes are
done by folding perturbative calculable partonic cross sections with
parton density functions that describe the structure of the hadrons as
given by the factorization theorem. The universality of the parton
density functions and the size of the expansion variable, the strong
coupling strength, $\alpha_s$ provide the basic parameters. A comparison
of these parameters for different processes and with other experiments
provides a basic means in testing QCD.

\section {Determination of the strong coupling strength}

\begin{figure}[tb]
 \begin{center}
  \includegraphics[width=.38\hsize]{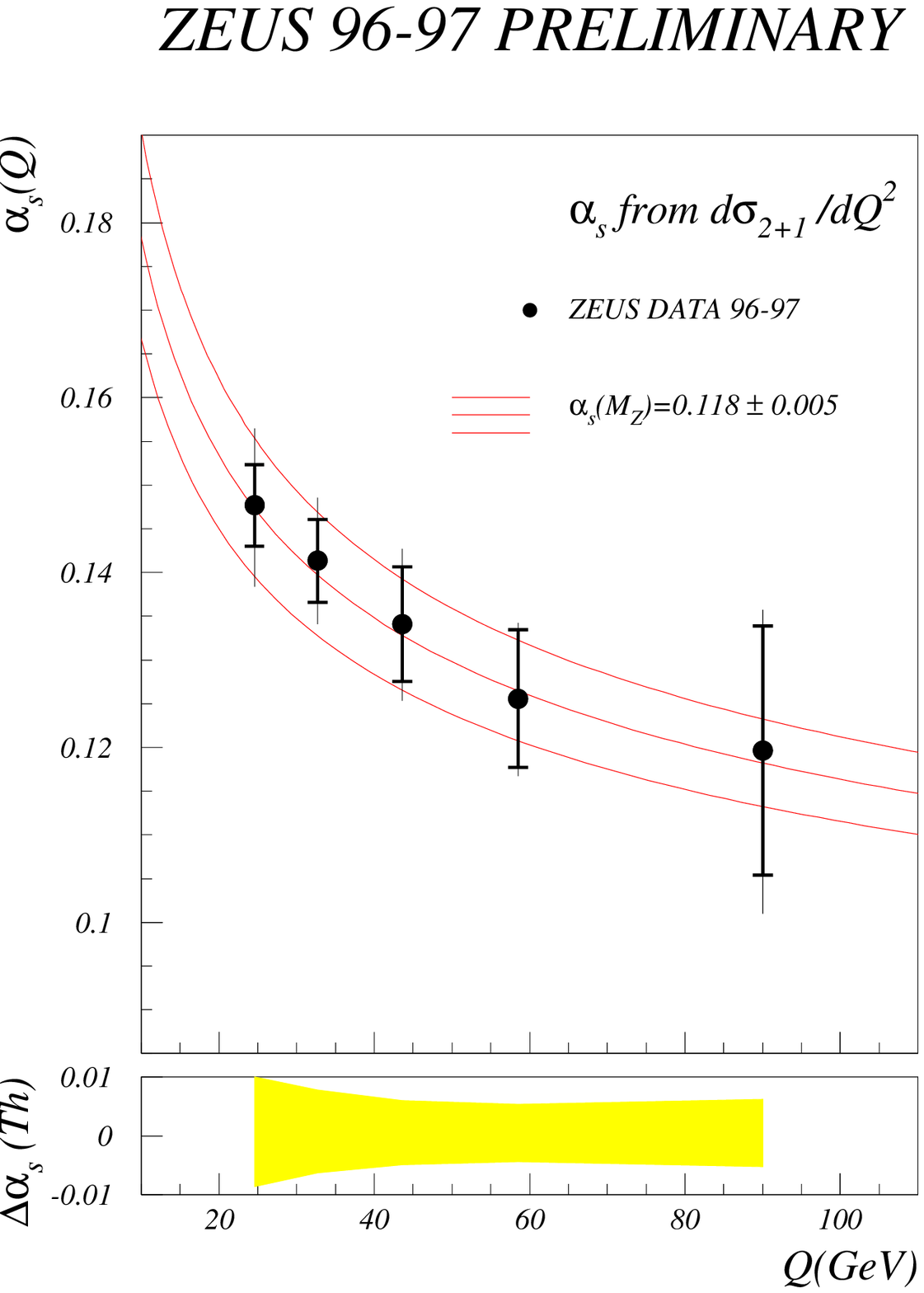}\hfil%
  \includegraphics[width=.38\hsize]{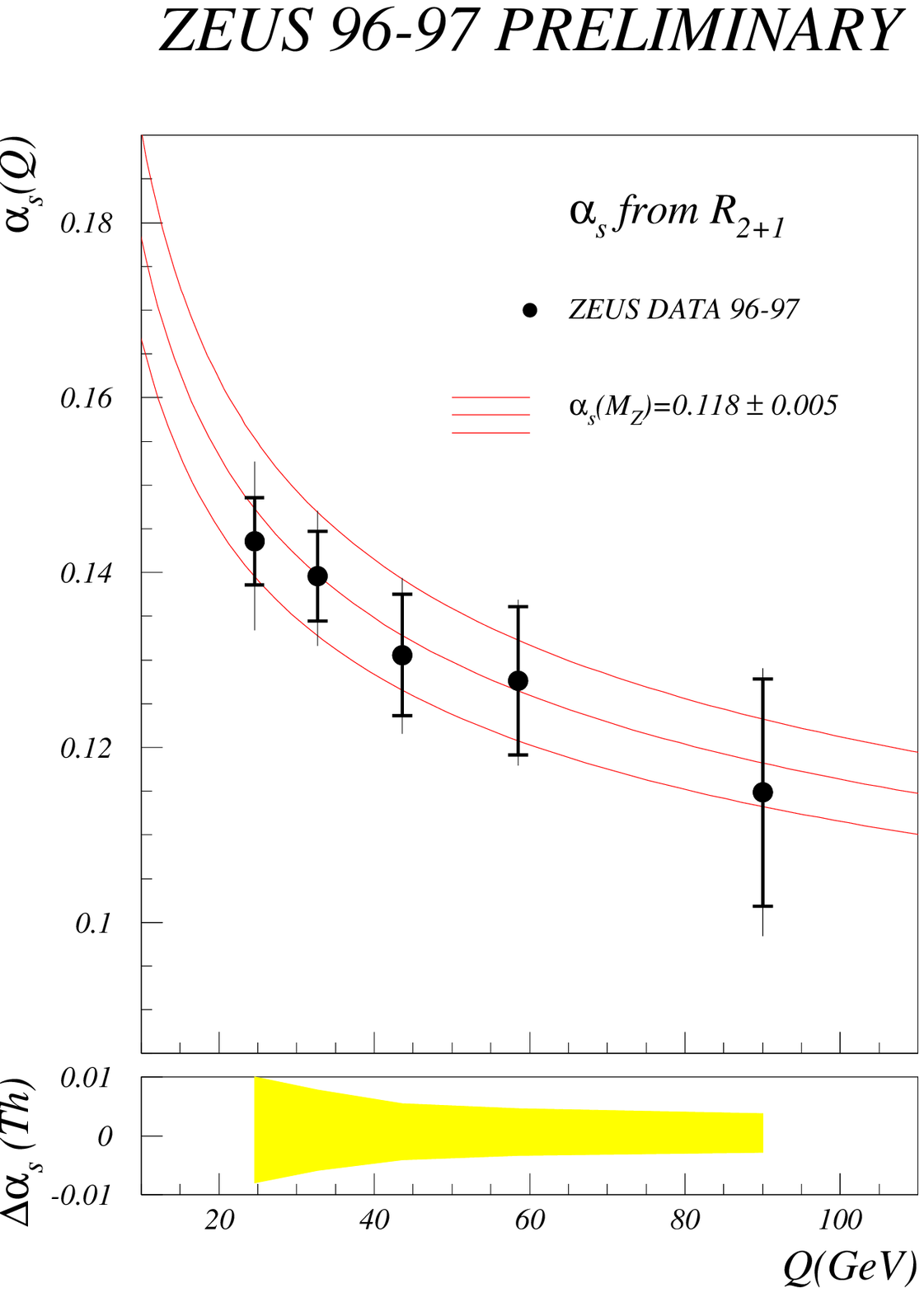}
 \end{center}
\caption{
\label{zeus:running:alphas}
The strong coupling strength $\alpha_s$ as a function of scale as
extracted by the ZEUS collaboration from dijet cross section (left) and
dijet rate (right). The data points are shown with the statistical error
(inner error bars) and the squared sum of the statistical and
experimental systematic error (full error bars). Additional correlated
theoretical errors shown in the band below the main plot are estimated
by varying the renormalization scale and the parton density functions.}
\end{figure}

The ZEUS collaboration has measured the inclusive and the dijet cross
section in bins of the virtuality $Q^2$ using the inclusive $k_t$
algorithm \cite{inclkt} in the data taken in 1996 and 1997.\cite{keil}
After correcting for detector and hadronization effects, NLO pQCD
calculations allow to extract the strong coupling strength $\alpha_s$
from the dijet cross section or the dijet rate. This is shown in
figure~\ref{zeus:running:alphas}.

A special emphasis was put on estimating the error induced by the parton
density functions ({\bf pdf}s) of the proton that are taken from other
experiments. Recently, a global fit by M.~Botje \cite{botje} was
published including information on the errors on the pdfs and their full
correlation.

A fit of all $\alpha_s(Q^2)$ values is performed and the result for the
strong coupling at the mass of the $Z$ boson extracted from the dijet
cross section is
\begin{eqnarray*}
\alpha_s(M_Z) &=& 0.1186 \pm 0.0019(\mbox{stat})
{+0.0020\atop -0.0007}(\mbox{exp})
{+0.0035\atop -0.0033}(\mbox{e.scale})\\
&&{+0.0048\atop -0.0038}(\mbox{ren.scale})
{\bf \pm 0.0031(\mbox{\bf pdf})}
\pm 0.0005(\mbox{hadr})
\end{eqnarray*}
and the one from the dijet rate is 
\begin{eqnarray*}
\alpha_s(M_Z) &=& 0.1166 \pm 0.0019(\mbox{stat})
{+0.0023\atop -0.0005}(\mbox{exp})
{+0.0036\atop -0.0034}(\mbox{e.scale})\\
&&{+0.0050\atop -0.0042}(\mbox{ren.scale})
{\bf {+0.0012\atop -0.0011}(\mbox{\bf pdf})}
\pm 0.0005(\mbox{hadr})
\end{eqnarray*}

It can be seen that the pdf uncertainties cancel partially for the rate.


\section {Determination of the gluon density}

Information on the gluon density in the proton can be extracted from two
sources: the scaling violations of the inclusive cross section and the
contribution of the boson gluon fusion process to the dijet cross
section. The H1 collaboration has measured both cross sections and
extracted the gluon density.\cite{dataincl,dijet}

A combined fit of both, the inclusive cross section $\sigma(x,Q^2)$ in
the region $20\GeV^2 \le Q^2 \le 5000\GeV^2$ and the dijet data
$\D{d^2\sigma_2\over dQ^2 dx}$ ($200\GeV^2 \le Q^2 \le 5000\GeV^2$) is
performed. While the inclusive data dominate the low $x$ region, the
dijet data contribute at medium $x$ ($0.01 \le x \le 0.1$) where data
from other processes are sparse.

\begin{figure}[tb]
 \begin{center}
  \includegraphics[width=.88\hsize]{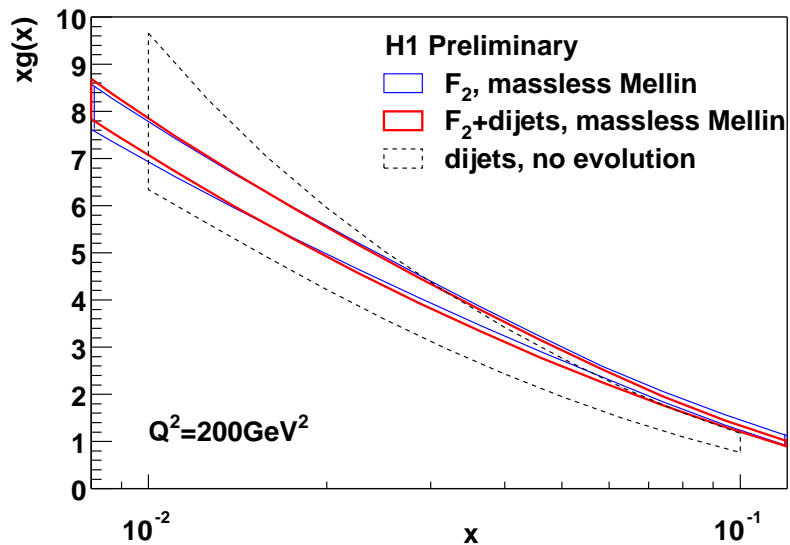}
 \end{center}
\caption{
\label{h1:gluon}
Gluon density at a scale of $200\GeV^2$ as a function of the fraction of
the parton momentum with respect to the incoming proton momentum as
extracted by H1. The thick, full line shows the gluon density of a fit
including dijet and inclusive data. For comparison fits with dijet
(dashed line) and inclusive data (thin, full line) are plotted.}
\end{figure}

Figure~\ref{h1:gluon} shows the gluon density in the medium $x$ region,
evolved in Mellin space to a scale of $200\GeV^2$ corresponding to the
average scale of the data included. The density extracted using the
information coming from the BGF process only has significantly larger
errors than the one extracted from scaling violations. In the combined
fit, both data sets are well described and the result is consistent with
the individual fits. This shows the independence of the pdfs of the
processes. However, adding the dijet information mainly influences the
fit stability and only a minor reduction in the uncertainty on the gluon
density is found.

\section {Conclusions}

The strong coupling strength has been extracted from dijet cross section
and dijet rate at HERA. In this analysis, the parton density functions
are taken from other measurements as determined in a global fit. The
values extracted from the dijet cross section and the dijet rate are
consistent with each other and with the current world average. The
determination has a precision that allows to have an impact on the world
average. For the first time, the uncertainty of the strong coupling due
to the uncertainty in the parton densities is estimated including the
full correlation between the different pdfs. It is demonstrated that
this uncertainty partially cancels when using the rate instead of the
cross section.

The gluon density is extracted for the first time using both, the size
of the scaling violations of the inclusive cross section and the direct
information from the dijet cross section. Both data sets, taken from
measurements of the H1 experiment only, are well described and the
resulting gluon density agrees with other extractions using a single
source of information only. The inclusion of the direct information from
the dijet cross section improves the fit stability compared to an
extraction using the inclusive cross section only.

The developments in the understanding of jet cross sections at the HERA
collider experiments demonstrate the importance of the lepton proton
scattering processes for testing QCD. The accuracy of the results is
comparable to those at other colliders and further precision tests can
expected for the new and upcoming data taking periods.


\bigskip

\section*{Acknowledgments}
I would like to thank the organizers of the DIS2000 conference for the
hospitality provided in Liverpool and the convenors of the WG 2 who
provided an interesting mix of talks on the different analysis testing
QCD.

\section*{References}



\end{document}